\documentclass[twocolumn,showpacs,preprintnumbers,amsmath,amssymb]{revtex4}

\usepackage{epsfig}
\usepackage{amsmath}
\usepackage{bm}
\usepackage{graphicx}

\newcommand{\ltsim}{\protect\raisebox{-0.5ex}{$\:\stackrel{\textstyle <}{\sim}\:$}}
\newcommand{\gtsim}{\protect\raisebox{-0.5ex}{$\:\stackrel{\textstyle >}{\sim}\:$}}

\begin{document}

\title{Temperature dependence of spinon and holon excitations in one-dimensional Mott insulators}
\author{H. Matsueda${}^{a}$}
\email{matsueda@imr.tohoku.ac.jp}
\author{N. Bulut${}^{a,b}$}
\author{T. Tohyama${}^{a}$}
\author{S. Maekawa${}^{a,b}$}
\affiliation{
${}^{a}$Institute for Materials Research, Tohoku University, Sendai 980-8577, Japan, \\
${}^{b}$CREST, Japan Science and Technology Agency (JST), Kawaguchi 332-0012, Japan}
\date{\today}
\begin{abstract}
Motivated by the recent angle-resolved photoemission spectroscopy (ARPES) measurements on one-dimensional Mott insulators, SrCuO${}_{2}$ and Na${}_{0.96}$V${}_{2}$O${}_{5}$, we examine the single-particle spectral weight of the one-dimensional (1D) Hubbard model at half-filling. We are particularly interested in the temperature dependence of the spinon and holon excitations. For this reason, we have performed the dynamical density matrix renormalization group and determinantal quantum Monte Carlo (QMC) calculations for the single-particle spectral weight of the 1D Hubbard model. In the QMC data, the spinon and holon branches become observable at temperatures where the short-range antiferromagnetic correlations develop. At these temperatures, the spinon branch grows rapidly. In the light of the numerical results, we discuss the spinon and holon branches observed by the ARPES experiments on SrCuO${}_{2}$. These numerical results are also in agreement with the temperature dependence of the ARPES results on Na${}_{0.96}$V${}_{2}$O${}_{5}$.
\end{abstract}
\maketitle

\section{Introduction}

One of the central issues in one-dimensional (1D) correlated electron systems is the spin-charge separation~\cite{Lieb}. In these systems, the spin and charge degrees of freedom of electrons are decoupled into their collective excitations, 'spinon' and 'holon'~\cite{TMO}. Since the single-particle excitations are not quasiparticles, it is expected that these excitations give rise to completely nontrivial physical properties.

Experimentally, angle-resolved photoemission spectroscopy (ARPES) gives direct information on the single-particle excitation spectra. High-quality ARPES measurements under various physical conditions enable us to understand the collective excitations in 1D compounds. Recently, such ARPES mearurements have been performed on a 1D Mott insulator SrCuO${}_{2}$, where the spinon and holon branches have been unambiguously observed~\cite{Kim}. In this work, the spectral weight from the main valence band with O$2p$ character is suppressed due to cross section effects, and thus the two branches become observable in the low binding-energy region. The band dispersions of the spinon and holon branches are in good agreement with those predicted in a spin-charge separated model~\cite{Kim2}. Furthermore, careful lineshape analysis reveals that the peak height of the holon branch is smaller than that of the spinon branch, and the full widths at half-maximum of the spinon and holon branches are estimated to be $\approx 0.7$ eV and $\approx 0.5$ eV, respectively.

Temperature dependent ARPES studies are also important, because finite-temperature effects on the single-particle excitation spectra of 1D Mott insulators are not due to simple thermal broadening. Since the single-particle excitation spectra are given by the convolution of the spinon and holon Green's functions, the finite-temperature effects reflect their collective nature, not the normal Fermi distribution of the quasiparticles. A temperature dependent ARPES study was carried out on Na${}_{0.96}$V${}_{2}$O${}_{5}$~\cite{Kobayashi}. In this work, the spectral weight redistribution from higher to lower binding-energy region was observed when the temperature was decreased from 300 K to 120 K. The redistribution occured on the scale of 1 eV, which was 100 times larger than the temperature change. The ARPES data were consistent with the finite-temperature exact-diagonalization calculations for the 1D $t$-$J$ model at half-filling, and it was shown that the spin-charge separation picture is valid for Na${}_{0.96}$V${}_{2}$O${}_{5}$.

In the high-energy ARPES measurements on SrCuO${}_{2}$~\cite{Kim}, it is quite significant to detect the spinon and holon branches directly in the ARPES spectrum, because the ARPES data enable us to examine the spectral weights and the lifetimes of the spinon and holon in the compound. The temperature dependent ARPES measurement on Na${}_{0.96}$V${}_{2}$O${}_{5}$ tells us how the spectral weights of the spinon and holon are redistributed with changing temperature, though the spinon and holon branches are not resolved in this experiment. The finite-temperature effect may also appear in the room-temperature ARPES data for SrCuO${}_{2}$.

Motivated by these ARPES measurements, we examine the single-particle excitation spectra in the 1D Hubbard model at half-filling. We are particularly interested in the temperature dependence of the spinon and holon excitations at temperatures of the order of the magnetic exchange $J\approx 4t^{2}/U$, where $t$ is the hopping matrix element and $U$ is the onsite Coulomb repulsion. In the large-$U$ limit, the single-particle spectral weight was obtained in Refs.~\cite{Ogata,Sorella,Parola,Penc,Penc2,Nakamura,Suzuura}. However, in the limit, the spinon excitations are not expected to exhibit temperature dependence. Therefore, such a simplified picture of the large-$U$ limit is not relevant for our discussion.

In order to study the temperature dependence, we calculate the single-particle spectral weight of the 1D half-filled Hubbard model by using the determinantal quantum Monte Carlo (QMC) and the maximum-entropy analytic continuation methods. For a complementary study to the QMC and a check of validity of our analytic continuation results, we also perform dynamical density matrix renormalization group (DDMRG) calculations for the spectral weight at zero temperature~\cite{White,Hallberg,Kuhner,Jeckelmann,Benthien}. It is noted that the single-particle spectrum of the 1D Hubbard model has been previously studied by using the QMC and the maximum-entropy techniques \cite{Preuss,Linden,Suga}. In Ref.~\cite{Preuss}, the single-particle spectrum and density of states were calculated for $U=4t$ and $T=0.0625t$ at half-filling and at $1/6$ doping. In addition, the velocities for the spin and charge excitations were obtained from the frequency and momentum dependences of the spin and charge susceptibilities. In Ref.~\cite{Suga}, the general features of the single-particle spectrum were discussed for $U=7.5t$ and $T=0.08t$ at half-filling. In these QMC data, however, the spinon and holon branches on the single-particle excitation spectrum were not resolved.

The organization of this paper is as follows. In the next section, we show the $U$ dependence of the spectral weight at zero temperature which was obtained by using the DDMRG method. In Section III, the temperature dependence of the spinon and holon excitations is discussed by combining the DDMRG and QMC results. The main purpose of this section is to resolve the spinon and holon excitations in the single-particle excitation spectrum of the 1D Hubbard model at half-filling using the QMC and the maximum-entropy techniques. The discussion and summary are given in Section IV.

\section{Zero-temperature dynamics}

The Hubbard Hamiltonian is given by
\begin{eqnarray}
H&=&-t\sum_{i,\sigma}( c^{\dagger}_{i,\sigma}c_{i+1,\sigma}+{\rm H.c.} )+U\sum_{i}n_{i,\uparrow}n_{i,\downarrow},
\end{eqnarray}
where $c_{i,\sigma}$ ($c_{i,\sigma}^{\dagger}$) annihilates (creates) an electron with spin $\sigma$ at lattice site $i$, $n_{i}=n_{i,\uparrow}+n_{i,\downarrow}$, $n_{i,\sigma}=c_{i,\sigma}^{\dagger}c_{i,\sigma}$, $t$ is the hopping integral along the chain axis, and $U$ is the on-site Coulomb repulsion. We consider the half-filled case.

In this section, we show the $U$ depencence of the spectral weights of the spinon and holon branches at zero temperature. The results will be helpful for understanding the $U$-dependence of the temperature evolution of the spinon and holon branches in the next section. The single-particle spectral weight is defined by
\begin{eqnarray}
A(k,\omega)=-\frac{1}{\pi}{\rm Im}\left<0\left|c^{\dagger}_{k,\uparrow}\frac{1}{E_{0}-\omega-H+i\gamma}c_{k,\uparrow}\right| 0\right> ,
\end{eqnarray}
where $c_{k,\uparrow}$ is the momentum representation of the electron operator with spin $\uparrow$, $\left|0\right>$ and $E_{0}$ are the ground state and the eigenenergy, respectively, and $\gamma$ is a small positive number.

In the DDMRG method, the superblock is set to be 64 sites, and we use the open boundary condition~\cite{White}. This is because the standard DMRG method gives the most precise numerical results when the open boundary condition is applied. However, there is a boundary effect, which makes a photohole localized at the ends of the 1D chain. For the single-particle spectral weight, it is not so easy to remove the boundary effect by increasing the system size. Thus, we add a potential of $-tn_{i}$ to the ends in order to remove the boundary effect. It is noted that the spinon branch is almost unchanged by the potential, while the holon branch is shifted to higher binding-energy region by the order of $0.1t$. The momentum representation of the electron operator, $c_{k,\uparrow}$, in the open boundary system is defined by
\begin{eqnarray}
c_{k,\uparrow}=\sqrt{\frac{2}{L+1}}\sum_{l=1}^{L}\sin(kl)c_{l,\uparrow} , 
\end{eqnarray}
in terms of an expansion in particle-in-a-box eigenstates, where $k=n\pi/(L+1)$ with $n=1,2,...,L$, and $L$ is the system size. In the infinite-chain limit, $L\rightarrow\infty$, this transformation becomes equivalent to the standard Fourier transformation. It may be technically useful to note how to count the fermion sign when the state $c_{k,\uparrow}\left|0\right>$ is generated, because the number of electrons at each site is neither 1 nor 0 in the DDMRG bases of $\left|0\right>$. Instead of $c_{l,\uparrow}$, we use an annihilation operator, $\tilde{c}_{l,\uparrow}$, which includes the fermion sign caused by the hopping from the $l$-th site to one end of the system (environment) block~\cite{Shibata}. The density matrix we introduce is composed of four target states, i.e., the ground state $\psi_{\alpha=1}=\left|0\right>$, the final state after annihilation of an electron $\psi_{\alpha=2}=c_{k,\uparrow}\left|0\right>$, and two correction vectors $\psi_{\alpha=3}=(E_{0}-\omega_{a}-H+i\gamma)^{-1}c_{k,\uparrow}\left| 0\right>$ and $\psi_{\alpha=4}=(E_{0}-\omega_{b}-H+i\gamma)^{-1}c_{k,\uparrow}\left| 0\right>$ being $\omega_{b}=\omega_{a}+2\gamma$. The method is the so-called  two correction vector method~\cite{Kuhner}. In the present work, the correction vectors are calculated by using the modified conjugate gradient method. By using these four target states, the density matrix $\rho$ for each $\omega$ is defined by
\begin{eqnarray}
\rho_{ii^{\prime}}(\omega)=\sum_{\alpha}p_{\alpha}\left[ \sum_{j}\psi_{\alpha,ij}^{\ast}\psi_{\alpha,i^{\prime}j} \right/ \left. \sum_{i,j}\psi_{\alpha,ij}^{\ast}\psi_{\alpha,ij} \right] ,
\end{eqnarray}
where the indices $i$ and $j$ run over all bases of the system and environment blocks, respectively, $\sum_{\alpha}p_{\alpha}=1$, and $p_{\alpha}$ is set to be $0.25$ for each $\alpha$. We have checked that the results are almost unchanged for other sets of $\left\{p_{\alpha}\right\}$. The DDMRG bases are truncated from the eigenstates of this density matrix. Our method is equivalent to the variational principle for the mixed states~\cite{White}. Since $\rho$ depends on $\omega$, the truncated bases are optimized independently for each $\omega$. This is the reason why the renormalization works well for dynamical quantities. The truncation number which is sufficient for convergence depends also on $\omega$. Here, the DDMRG bases are truncated up to $m=400$. After convergence of the finite-system DMRG algorithm, we obtain $A(k,\omega_{a})=-{\rm Im}\left<\psi_{2}|\psi_{3}\right>/\pi$ and $A(k,\omega_{b})=-{\rm Im}\left<\psi_{2}|\psi_{4}\right>/\pi$. At the same time, the spectral weights at an energy interval $\omega_{a}<\omega<\omega_{b}$ are calculated by using the same $E_{0}$ and $H$. The broadening factor is set to be $\gamma=0.1t$, which is less than temperatures treated in the next section. It is noted that the two correction vector method is efficient, but we need a higher truncation number than that in the standard DDMRG with a single correction vector. Then, we check the convergency for some $\omega$ points by using three target states with the single correction vector and by taking the truncation number up to $m=480$.

\begin{figure}
\begin{center}
\epsfig{file=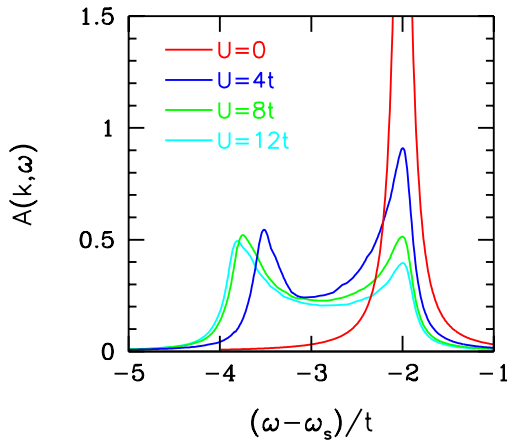,height=6cm}
\end{center}
\caption{ (Color online) $A(k, \omega)$ versus $\omega$ for various values of $U/t$ obtained by the DDMRG method. The momentum $k$ is set to be $\pi/65$, which is the smallest value in the DDMRG calculations. The red line is the quasiparticle peak for $U/t=0$, obtained for $k=0$ by using periodic boundary conditions and a finite broadening of $0.1t$. The origin of frequency is shifted by $\omega_{s}$ for the finite-$U$ values, so that all of the peak positions occur at $\omega=-2t$. The value of the shift $\omega_{s}$ is $\approx 0.0$, $-1.13t$, and $-2.80t$ for $U=4t$, $8t$, and $12t$, respectively.}
\label{fig1}
\end{figure}

In Fig.~\ref{fig1}, we show the $U$ dependence of the spectral weight at $k=\pi/65$, which is the smallest momentum in the open boundary system. In finite-$U$ cases, we find two peaks at the band edges. The peaks at low and high binding-energy sides correspond to the spinon and holon excitations, respectively, because both peak positions are equal to those predicted by the Bethe-anzats solutions. In this figure, these peaks do no show clear branch cuts due to the finite broadening factor of $\gamma=0.1t$. However, we have checked that the edge singularities at the peak positions become observable with smaller values of $\gamma$.

In the case of $U/t\ltsim 8$ ($U/t\gtsim 8$), the peak height of the spinon (holon) branch is larger than that of the holon (spinon) branch. In the finite-$U$ cases, the spectral shape of the holon branch is almost unchanged, while the weight of the spinon branch rapidly decreases with decreasing temperature.

\section{The spectral weight at finite temperatures}

Here, we discuss the temperature dependence of $A(k,\omega)$ of the 1D Hubbard model. For this purpose, we present data obtained by using the determinantal QMC technique \cite{White2}. With this method, we have calculated the single-particle Green's function
\begin{equation}
G(k,\tau) = - \sum_{\ell} e^{-i k r_{\ell}} \langle T_{\tau} \,
c_{i+\ell,\sigma}(\tau) c_{i,\sigma}^{\dagger}(0) \rangle,
\end{equation}
where $\left<\cdot\cdot\cdot\right>$ denotes thermal averaging, $c_{i\sigma}(\tau) = e^{H\tau} c_{i\sigma} e^{-H\tau}$, and $T_{\tau}$ is the Matsubara time-ordering operator. For temperature $T$ (inverse temperature $\beta$), the integral equation
\begin{equation}
G(k,\tau) = \int_{-\infty}^{+\infty} d\omega \, {e^{-\tau\omega}
\over 1 + e^{-\beta\omega} } A(k,\omega)  \label{Gpt}
\end{equation}
expresses $G(k,\tau)$ in terms of the single-particle spectral weight
\begin{equation}
A(k,\omega)=-{1\over \pi}{\rm Im}\,G(k,i\omega_n \rightarrow
\omega+i\gamma)
\end{equation}
where
\begin{equation}
G(k,i\omega_n) = \int_0^{\beta} d\tau\, e^{i\omega_n\tau}
G(k,\tau)
\end{equation}
and $\omega_n=(2n+1)\pi T$ is the fermion Matsubara frequency. We have obtained $A(k,\omega)$ from the QMC data on $G(k,\tau)$ by solving Eq.~(\ref{Gpt}) with the maximum-entropy analytic continuation method.

We also present results on the static magnetic susceptibility at zero frequency, $\chi(q)$, which is defined by
\begin{equation}
\chi(q) = \int_0^{\beta} d\tau \sum_{\ell} e^{-iq\ell} \langle m^z_{i+\ell}(\tau) m^z_i(0) \rangle,
\end{equation}
where $m^z_i = n_{i\uparrow}-n_{i\downarrow}$ and $m^z_i(\tau) = e^{H\tau} m^z_i e^{-H\tau}$.

The determinantal QMC technique does not have the "fermion sign problem" \cite{Loh} at half-filling or in the 1D case. However, the QMC algorithm which proceeds by using local updates of the Hubbard-Stratonovich spins produces results with long autocorrelation times at low temperatures and for large values of $U/t$ \cite{Scalettar}.  It has been noted that this is because the determinantal QMC algorithm with single-spin-flip moves does not explore the phase space in an ergodic manner for large values of $U/t$. In order to remove this problem, global Monte Carlo moves have been introduced, and this way the antiferromagnetic structure factor for the $4\times 4$ Hubbard lattice was calculated for $U$ up to $16t$ \cite{Scalettar}. However, the effects of global moves on dynamical quantities have not been explored. Because of these reasons, we have restricted our QMC calculations, which use single-spin-flip moves, to parameter regimes where we did not observe ergodicity problems. In the following, we present QMC data on $A(k,\omega)$ for $U=4t$ at $0.25t\le T\le 1.0t$ and for $U=8t$ at $0.33t \le T\le 1.0t$, and make comparisons with the DDMRG data obtained at $T=0$.
   
In obtaining $A(k,\omega)$ from QMC data on $G(k,\tau)$, we have used the maximum-entropy analytic continuation procedure described in Ref. \cite{Jarrell}. As it is well known, the maximum entropy technique has finite resolution which decreases away from the Fermi level. Furthermore, at low temperatures and large values of $U/t$, the $G(k,\tau)$ data exhibit long autocorrelation times \cite{Scalettar}. In order to improve the accuracy of the maximum-entropy results for $A(k,\omega)$, we have obtained QMC data on $G(k,\tau)$ with good statistics. For example, the
covariance matrix of $G(k,\tau)$ used in the maximum-entropy technique always exhibited a continuous eigenvalue spectrum, as discussed by Ref.~\cite{Jarrell}. In addition, the Bryan's and the classical maximum-entropy algorithms \cite{Jarrell} produced similar results for $A(k,\omega)$. Furthermore, we have monitored the results of the maximum-entropy procedure for $A(k,\omega)$ as the statistics of the QMC data on $G(k,\tau)$ improved. These provide information about the reliability of the maximum-entropy images of $A(k,\omega)$ presented in this section.

In the following, we present QMC data for a 32-site chain with periodic boundary conditions, and make comparisons with the DDMRG data obtained for a 64-site chain. Here, $A(k,\omega)$ is plotted in units of $t^{-1}$, and $A(k,\omega)=A(\pi-k,-\omega)$ at half-filling. It is noted that the momenta which we can access in the DDMRG method are different from those in the QMC method, because different boundary conditions are taken into account in these methods in order to keep numerical precision. Then, the QMC result with momentum $k$ is compared with the DDMRG one with momentum $Lk/(L+1)$, and the QMC result with $k=0$ is also compared with the DDMRG one with the smallest momentum $\pi/(L+1)$. In order to make the difference small, we take the system size $L$ as large as possible in the DDMRG method.

\begin{figure}
\begin{center}
\epsfig{file=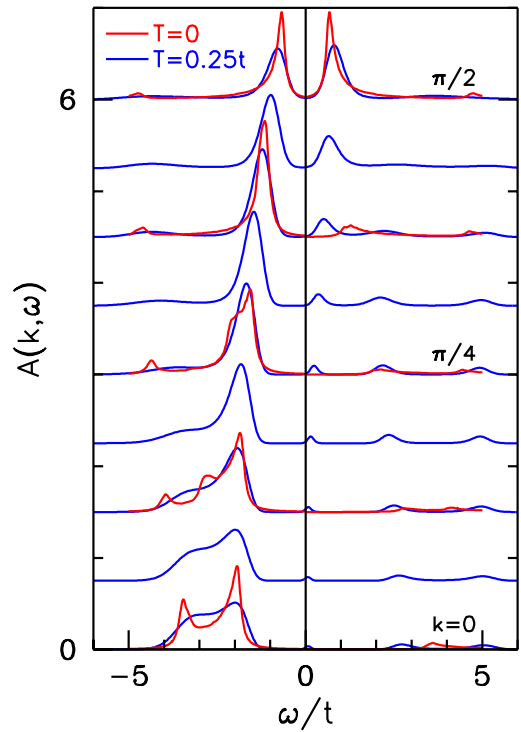,height=10cm}
\end{center}
\caption{ (Color online) Single-particle spectral weight $A(k,\omega)$ for the 1D Hubbard model at half-filling for $U=4t$. The blue curves denote the QMC data at $T=0.25t$ for wavevectors $0\le k\le \pi/2$ for a 32-site chain with periodic boundary conditions. The red curves denote the DDMRG data at $T=0$ for wavevectors $k=0$, $\pi/8$, $\pi/4$, $3\pi/8$ and $\pi/2$ for a 64-site chain with open boundary conditions. For definition of $k$ in the DDMRG data, see the text.}
\label{fig2}
\end{figure}

Figure~\ref{fig2} shows the QMC and DDMRG results on $A(k,\omega)$ for $U=4t$ at half-filling. Here, the QMC results were obtained at $T=0.25t$, and the DDMRG data are at $T=0$. For $k=0$, the DDMRG data show that the holon and spinon branches of $A(k,\omega)$ are at $\omega\approx-3.5t$ and $\approx -2t$, respectively. In addition, the insulating gap and small amount of spectral weight at $\omega\approx 3.5t$ are observed. For $k=\pi/2$, the peak in the
DDMRG results for $A(k,\omega)$ occurs at $\approx -0.7t$. Hence, we deduce that $\Delta\approx 0.7t$ for the Mott-Hubbard gap. As we go from $k=0$ to $\pi/8$, the DDMRG results show that the spinon peak exhibits weak dispersion, while the holon peak moves
rapidly towards the spinon branch. As $k\rightarrow \pi/2$, the spinon and the holon branches merge together. At $T=0.25t$ and for $k=0$, the maximum-entropy images of $A(k,\omega)$ exhibit a peak at $\omega\approx -2t$ and a shoulder centered at $\approx -3.2t$, which we attribute to the spinon and holon excitations, respectively. However, it is not possible to resolve the holon and spinon branches for $k=\pi/8$ at $T=0.25t$. We note that, at $T=0.25t$, as $k$ goes from $3\pi/8$ to $\pi/2$, the height of the peak in $A(k,\omega)$ decreases. This behavior is also observed for the DDMRG data at $T=0$. For $k=\pi/2$, the peak in $A(k,\omega)$ at $T=0.25t$ is significantly rounded compared to the DDMRG data at $T=0$, which includes an artificial broadening of $\gamma=0.1t$. We attribute the rounding of the peak in the QMC data for $A(k=\pi/2,\omega)$ to finite-temperature effects. For $k=3\pi/8$, we also observe that,
at $T=0.25t$, there are single-particle excitations at $\omega\approx 0$ of which intensity decreases as $k$ goes towards the zone center.

\begin{figure}
\begin{center}
\epsfig{file=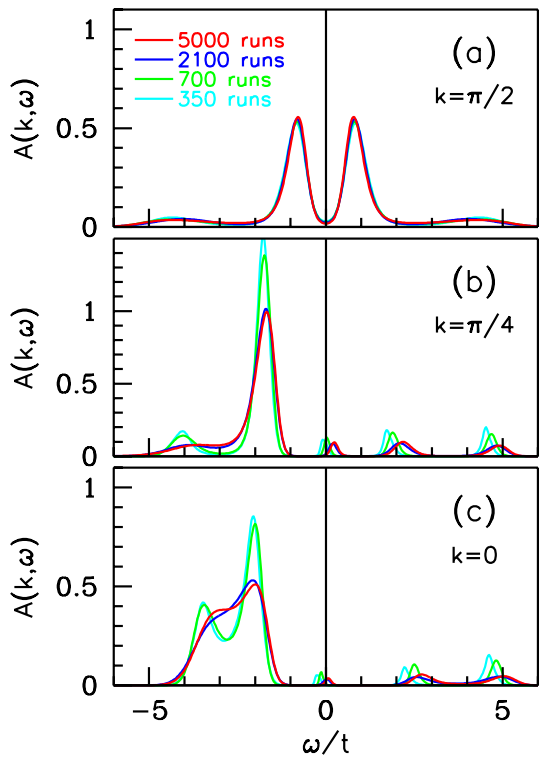,height=10cm}
\end{center}
\caption{ (Color online) Evolution of the maximum-entropy images for $A(k,\omega)$ as the statistics of the QMC data on $G(k,\tau)$ improves. Here, $A(k,\omega)$ is plotted as the number of runs used for obtaining $G(k,\tau)$ increases from 350 to 5000, where one run corresponds to one independent QMC simulation with $4\times 10^5$ Monte Carlo sweeps. These results are shown for wavevectors $k=\pi/2$, $\pi/4$ and $0$. }
\label{fig3}
\end{figure}

Figure~\ref{fig3} shows the evolution of the maximum-entropy images of $A(k,\omega)$ as the statistics of $G(k,\tau)$ improve for $U=4t$ and $T=0.25t$. This is useful in order to have an estimate of the resolution of the maximum-entropy method. We note that the dependence of the maximum-entropy results for $A(k,\omega)$ on the statistics of the QMC data was also investigated in
Ref.~\cite{Linden} for the 12-site Hubbard chain with $U=4t$ and $T=0.05t$. In Fig.~\ref{fig3}, $A(k,\omega)$ versus $\omega$ is plotted for $k=\pi/2$, $\pi/4$ and 0, as the number of runs increases from 350 to 5000. Here, each run corresponds to one independent QMC simulation with $4\times 10^5$ updates of the Hubbard-Stratonovich fields. We find that, for $k=\pi/2$, the maximum-entropy results on $A(k,\omega)$ converge rapidly. However, for $k=0$, the resolution worsens, in particular, for structures in $A(k,\omega)$ away from the Fermi level. Hence, it is necessary to have QMC data with good statistics in order to resolve reliably the features away from the Fermi level. Here, we observe that the maximum-entropy images of $A(k=0,\omega)$ for 2100 and 5000 runs are similar. The QMC results on $A(k,\omega)$ were obtained by performing approximately 5000 runs.

\begin{figure}
\begin{center}
\epsfig{file=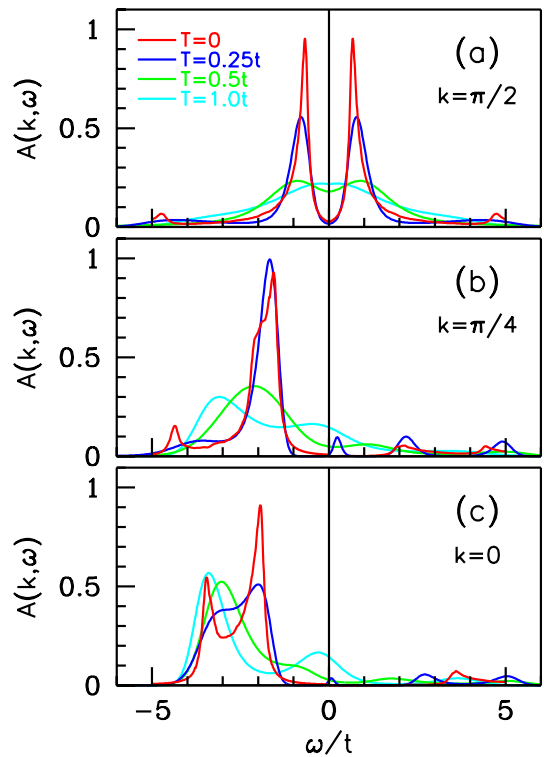,height=10cm}
\end{center}
\caption{ (Color online) Temperature dependence of $A(k,\omega)$ versus $\omega$ for $U=4t$ and half-filling for wavevectors $k=\pi/2$, $\pi/4$ and 0. Here, the finite-temperature results are from maximum-entropy analytic continuation of QMC data for a 32-site chain with periodic boundary conditions, and the $T=0$ results were obtained by DDMRG for a 64-site chain with open boundary conditions.}
\label{fig4}
\end{figure}

In Fig.~\ref{fig4}, we show results on the $T$ dependence of $A(k,\omega)$ for $U=4t$ at half-filling for wavevectors $k=0$, $\pi/4$ and $\pi/2$. At $T=1.0t\gtsim \Delta$, $A(k=\pi/2,\omega)$ exhibits a broad peak centered at $\omega=0$, while at $T=0.5t\ltsim\Delta$ the insulating gap starts to develop. For $k=0$, the $T$ dependence is more involved. Here, at $T=1.0t\gtsim\Delta$, a significant amount of the spectral weight is observed at $\omega\approx -3.5t$, which corresponds to the location of the holon peak at $T=0$. Furthermore, an additional peak at $\omega\approx 0$ and also spectral weight at $\omega > 0$ are observed. At $T=1.0t\gtsim\Delta$, there is a pseudogap at $\omega\approx -2t$, which is the location of the quasiparticle peak for the noninteracting system. As $T$ decreases from $1.0t$ to 0, we observe that spectral weight from $\omega\approx -3.5t$ and from $\omega\approx 0$ are transferred to $\omega\approx -2t$ to form the spinon peak. We think that the pseudogap at $\omega\approx -2t$ is due to the onsite Coulomb repulsion and, for $T \gtrsim U$,  the peak in $A(k=0,\omega)$ occurs at $\omega=-2t$. However, we also observe that, for $k=0$, there is significant amount of spectral weight at the frequency of the holon excitations already for $T \lesssim U$, while the spinon branch develops for $T\lesssim 1.0t \approx J$. The temperature evolution for $k=\pi/4$ is similar to that for $k=0$. However, the spinon and holon peaks are located closer, and it is not possible to resolve them at $T=0.25t$.

\begin{figure}
\begin{center}
\epsfig{file=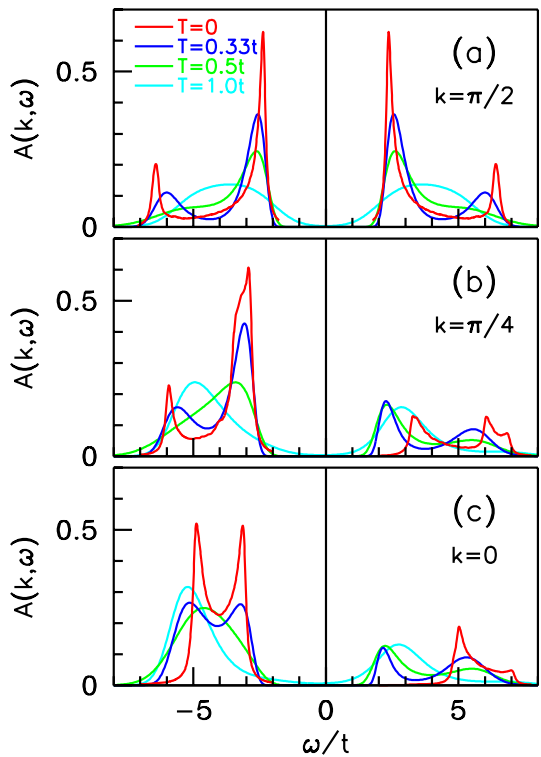,height=10cm}
\end{center}
\caption{ (Color online) Temperature dependence of $A(k,\omega)$ versus $\omega$ for $U=8t$ and half-filling, plotted in the same way as in Fig.~4. }
\label{fig5}
\end{figure}

In Fig.~\ref{fig5}, we show results on the $T$ dependence of $A(k,\omega)$ for $U=8t$ at half-filling. For $k=\pi/2$, we observe that the Mott-Hubbard gap is approximately $2.4t$. 
In this case, we have the magnetic exchange $J\approx 0.5t$, and we observe that $A(k,\omega)$ exhibits strong $T$-dependence. In particular, at $T=0.33t\ltsim J$ and for $k=0$, the maximum-entropy image of $A(k,\omega)$ shows a double peak structure of almost the same peak heights. The peak positions are nearly equal to those at $T=0$, and thus the peaks at $T=0.33t$ are attributed to the spinon and holon branches. The spinon branch becomes observable at $T=0.33t$.

\begin{figure}
\begin{center}
\epsfig{file=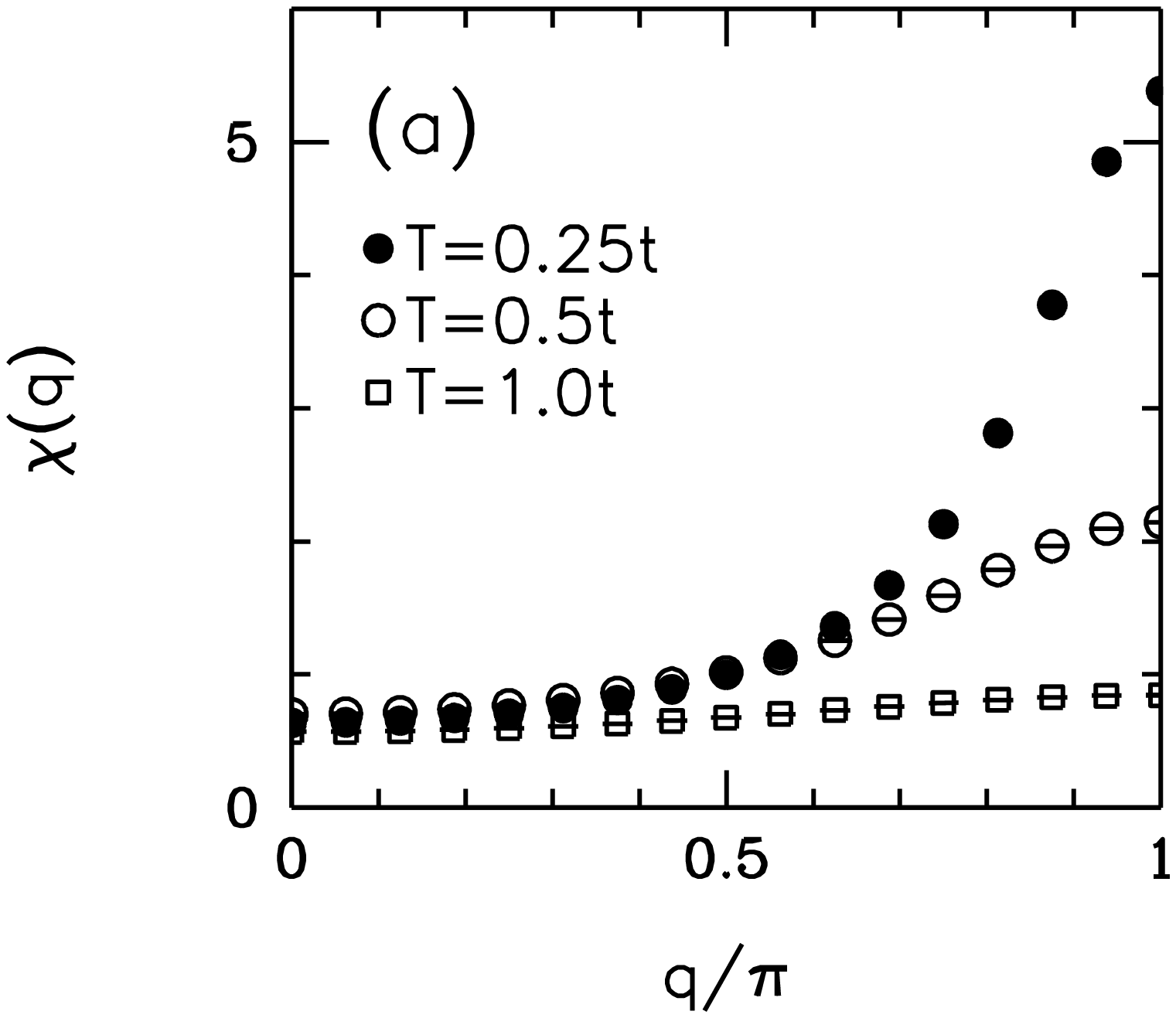,height=6cm}
\epsfig{file=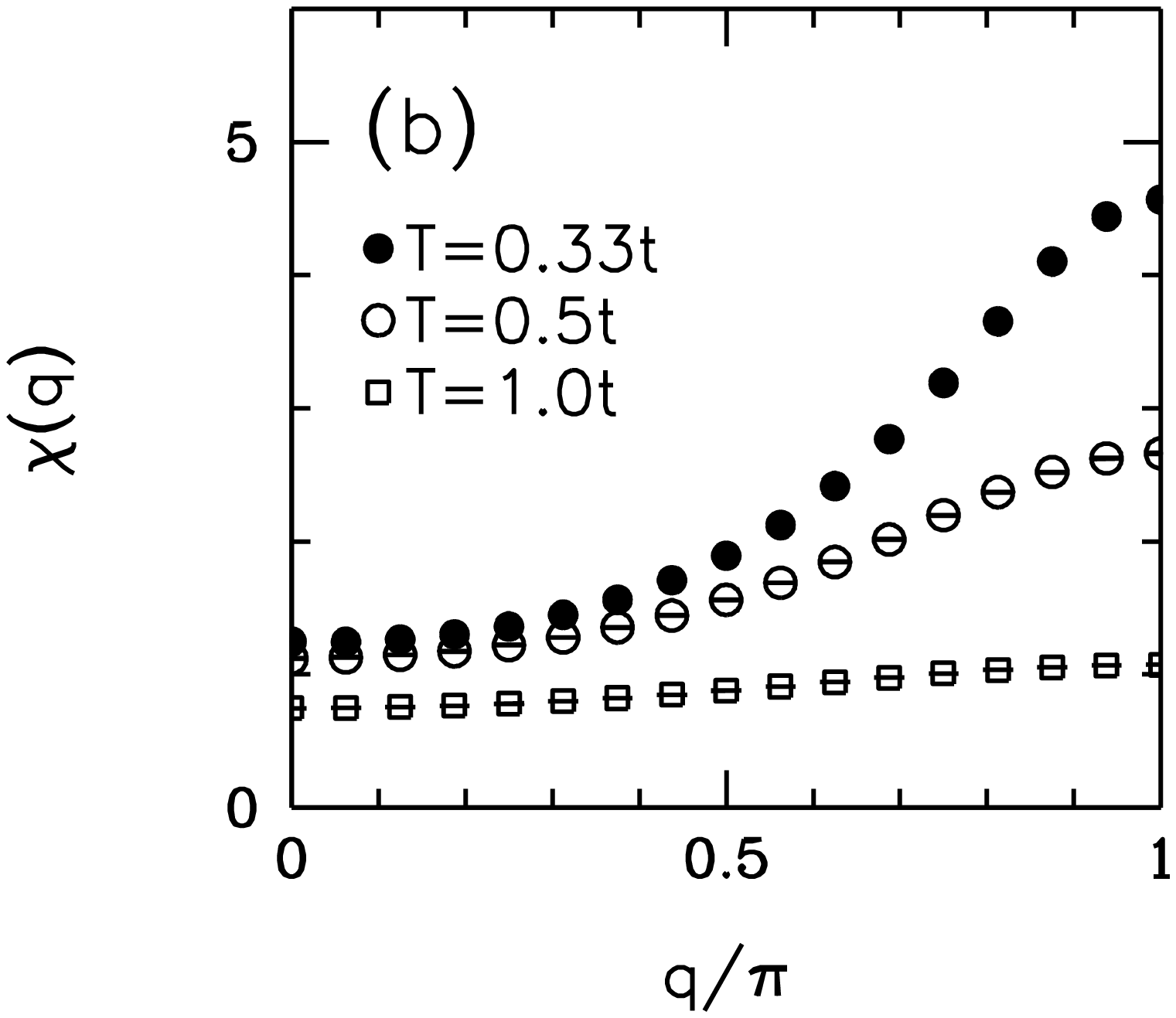,height=6cm}
\end{center}
\caption{ Magnetic susceptibility of the 1D Hubbard model at zero frequency $\chi(q)$ versus $q$ for (a) $U=4t$ and (b) $U=8t$. Here, results are shown at half-filling for various temperatures.}
\label{fig6}
\end{figure}

\begin{figure}
\begin{center}
\epsfig{file=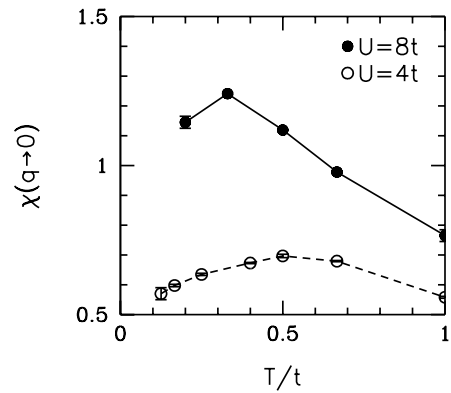,height=6cm}
\end{center}
\caption{ Temperature dependence of the uniform magnetic susceptibility $\chi(q\rightarrow 0)$ for $U=4t$ and $8t$ at half-filling. }
\label{fig7}
\end{figure}

Finally, we show the magnetic susceptibility data. Figures~\ref{fig6}(a) and (b) show QMC results on the temperature dependence of the magnetic susceptibility $\chi(q)$ for $U=4t$ and $8t$, respectively. In these figures, $\chi(q)$ versus $q$ is plotted for the same temperatures as those used in Figs.~\ref{fig4} and~\ref{fig5}. For these $U$ values, $\chi(q\rightarrow\pi)$ increases rapidly as $T$ decreases. Figure~\ref{fig7} shows QMC results on the temperature dependence of the uniform susceptibility $\chi(q\rightarrow 0)$ for $U=4t$ and $8t$. By determining the maximum of $\chi(q\rightarrow 0)$, we estimate the temperature where the antiferromagnetic correlations develop. The temperature is estimated to be $T\approx 0.5t$ for $U=4t$, and $T\approx 0.33t$ for $U=8t$. For $U=8t$, the value $0.33t$ is close to $J\approx 0.5t$, while, for $U=4t$, the value $0.5t$ is far below $J\approx 1.0t$.

\section{Summary and Discussion}

Here, we summarize the numerical results obtained in the previous two sections, and discuss recent experiments in the light of them. In Section II, the $U$ dependence of $A(k\approx 0,\omega)$ was calculated by the DDMRG method at $T=0$. For the values of $U$ presented in Fig. 1 ($4t\le U\le 12t$), the spinon and holon branches are clearly resolved for $k\approx 0$. We have also checked that the spin-charge separation is observable for $U$ down to $1.0t$, though we do not show these data. In Fig.~\ref{fig1}, the line shape of the holon branch is almost unchanged for $4t\le U\le 12t$, while the peak height of the spinon branch rapidly increases as $U$ decreases from $12t$ to $4t$.

In Section III, the $T$ dependence of $A(k,\omega)$ was presented at half-filling in the QMC method. $A(k\rightarrow 0,\omega)$ clearly shows the double peak structure at $T=0.25t$ and $0.33t$ for $U=4t$ and $8t$, respectively. At these temperatures, the uniform magnetic susceptibility has a maximum as a function of $T$, and the short-range antiferromagnetic correlations develop.

Let us first compare the numerical results with the ARPES data for Na${}_{0.96}$V${}_{2}$O${}_{5}$~\cite{Kobayashi}. The value of $U$ of this compound is calculated to be $U\approx 12t$ in the finite-temperature exact-diagonalization calculations for the 1D $t$-$J$ model~\cite{Kobayashi}. Therefore, the results for $U=8t$ may be applicable for comparisions at a qualitative level. The DDMRG (QMC) results are obtained at $T=0$ ($0.33t$), whereas the temperature region where the ARPES measurements were carried out is estimated to be between $0$ and $0.33t$. In the numerical results on $A(k,\omega)$ for $k=0$, $\pi/4$ and $\pi/2$ shown in Fig.~\ref{fig5}, the peak height of the spinon branch increases as $T$ decreases from $0.33t$ to $0$. In this temperature region, the threshold at low energy side is almost unchanged. For $k=0$, the width of the lower Hubbard band becomes narrower with decreasing $T$, while for $k=\pi/2$, the width is almost independent of $T$ for $T\le 0.5t$. These results are consistent with the ARPES data on Na${}_{0.96}$V${}_{2}$O${}_{5}$.

Next, we compare the results for $U=8t$ with the ARPES data for SrCuO${}_{2}$~\cite{Kim}, since the appropriate value of $U$ for this compound is estimated to be $U\approx 10t$. In the ARPES data obtained at room temperature, the spinon and holon branches are resolved near $k=0$. As $k$ goes from $0$ to $\pi/2$, the spinon and holon branches merge into a single peak. The band widths of the spinon and holon branches are consistent with the numerical results shown in Fig. 5. The peak height of the holon branch is smaller than that of the spinon. On the pther hand, in the numerical results for $A(k\rightarrow 0,\omega)$ at $T=0$ and $0.33t$ shown in Fig.~\ref{fig5}, the spinon and holon branches have almost equal spectral weights. The discrepancy between the theory and the experiment may be attributed to the electron-phonon interaction. Within this context, it has been shown that the electron-phonon interaction broadens the holon branch more than the spinon branch~\cite{Tsutsui}. The detailed origin of this remains to be clarified.

\section{Acknowledgments}

We acknowledge K. Tsutsui for the comments based on the exact diagonalization results for the Hubbard model. This work was supported by NAREGI Nanoscience Project and Grant-in Aid for Scientific Research from the Misistry of Education, Culture, Sports, Science and Technology of Japan, and NEDO. One of us (N.B.) would like to thank the International Frontier Center for Advanced Materials at Tohoku University for its kind hospitality, and gratefully acknowledges support from the Japan Society for the Promotion of Science (JSPS) through the JSPS fellowship and from the Turkish Academy of Sciences through the GEBIP program (EA-TUBA-GEBIP/2001-1-1).

\end{document}